\documentclass{optica-article}

\usepackage[english]{babel}	
\usepackage{physics}	
\usepackage{mathtools}	
\usepackage{nicefrac}		
\usepackage{xcolor}	
\usepackage{algorithm}
\usepackage{enumitem} 
\usepackage{appendix}	

\graphicspath{{images/}}

\newcommand{\CCC}{\mathcal{C}}
\newcommand{\PPP}{\mathcal{P}}
\newcommand{\RR}{\mathbb{R}}
\newcommand{\NN}{\mathbb{N}}

\renewcommand{\pr}{\prime}

\newcommand{\eps}{\varepsilon}
\newcommand{\mr}[1]{\mathrm {#1}}
\newcommand{\Vsup}{{V_\mathrm{S}}}
\newcommand{\Vmode}{{V_\mathrm{M}}}
\newcommand{\ie}{\textit{i.e.}}
\newcommand{\eg}{\textit{e.g.}}

\renewcommand{\dd}{\mathrm{d}}
\newcommand{\Vcav}{V_\mathrm{C}}

\newcommand{\Esup}{E_\mathrm{S}}

\newcommand{\eq}[1]{(\ref{#1})}

\newcommand{\VV}{\widetilde V}
\newcommand{\EE}{\widetilde E}
\newcommand{\XX}{\mathcal{X}}
\newcommand{\eqdef}{\coloneqq}
\newcommand{\rr}{{\vb r}}
\newcommand{\ra}{\rightarrow}

\journal{oe}

\articletype{Research Article}

\begin{document}

\title{Unsupervised Machine Learning to Classify the Confinement of Waves in Periodic Superstructures} 

\author{Marek Kozo\v{n},\authormark{1,2,*} Rutger Schrijver,\authormark{2} Matthias Schlottbom,\authormark{2} Jaap J.W. van der Vegt,\authormark{2} and Willem L. Vos\authormark{1}}

\address{\authormark{1}Complex Photonic Systems (COPS), MESA+ Institute for Nanotechnology, University of Twente, P.O. Box 217, 7500 AE Enschede, The Netherlands\\
\authormark{2}Mathematics of Computational Science (MACS),  MESA+ Institute for Nanotechnology, University of Twente, P.O. Box 217, 7500 AE Enschede, The Netherlands}

\email{\authormark{*}m.kozon@utwente.nl} 



\begin{abstract*}
We employ unsupervised machine learning to enhance the accuracy of our recently presented scaling method for wave confinement analysis~\cite{Kozon2022Phys.Rev.Lett.}.
We employ the standard k-means++ algorithm as well as our own model-based algorithm.
We investigate cluster validity indices as a means to find the correct number of confinement dimensionalities to be used as an input to the clustering algorithms.
Subsequently, we analyze the performance of the two clustering algorithms when compared to the direct application of the scaling method without clustering.
We find that the clustering approach provides more physically meaningful results, but may struggle with identifying the correct set of confinement dimensionalities.
We conclude that the most accurate outcome is obtained by first applying the direct scaling to find the correct set of confinement dimensionalities and subsequently employing clustering to refine the results. 
Moreover, our model-based algorithm outperforms the standard k-means++ clustering.
\end{abstract*}

\section{Introduction} 
Completely controlling wave propagation in periodic media is a key challenge that is essential for a large variety of applications ~\cite{Markos2008, Fink2000Rep.Prog.Phys., Liu2000Science, Maldovan2013Nature, Cummer2016Nat.Rev.Mater., Kruglyak2010J.Phys.D, Wagner2016Nat.Nanotechnol., Klyukin2018Phys.Rev.Lett., Callahan2013Opt.Express, Tandaechanurat2011Nat.Photonics, Aspelmeyer2014Rev.Mod.Phys., Koenderink2015Science, Li2018Opt.Express, Wang2020Nat.Photonics, Uppu2021Phys.Rev.Lett.}. 
An especially interesting type of control is wave confinement achieved by introducing disorder and functional defects into an otherwise periodic medium~\cite{Villeneuve1996Phys.Rev.B,Anderson1958Phys.Rev.,Koenderink2005Phys.Rev.B,Conti2008NaturePhys.}. 
The interference of waves in such an altered structure may result in a strong concentration of the energy density inside a small sub-volume of the medium.
Wave confinement has been investigated for different types of waves and in various settings, \textit{e.g.}, classical mechanics~\cite{Arceri2020Phys.Rev.Lett.}, photonics~\cite{Callahan2013Opt.Express, Tandaechanurat2011Nat.Photonics, Busch2007Phys.Rep., Woldering2014Phys.Rev.B, Hack2019Phys.Rev.B}, solid state physics~\cite{Economou2010,Shao2008J.Phys.Chem.C, Pashartis2017Phys.Rev.Appl., Pashartis2017Phys.Rev.B, Zhang2011Phys.Rev.B}, or magnonics~\cite{Demokritov2017book, Tartakovskaya2016Phys.Rev.B}.
Its applications include sensors, controlled spontaneous emission, and enhanced interactions between hybrid wave-types such as sound and light~\cite{Krioukov2002Opt.Lett.,Baba2008Nat.Photonics, Noda2000Nature,Gerard1998Phys.Rev.Lett.,Michler2003,Reithmaier2004Nature,Yoshie2004Nature,Peter2005Phys.Rev.Lett.,Russell2003Opt.Express}.

Recently, we have described a rigorous method to characterize the confinement of waves in periodic media with defects and in superlattices in general~\cite{Kozon2022Phys.Rev.Lett.}. 
We first introduced a so-called confinement dimensionality, that quantifies the intuitive term of "confinement" and then developed a scaling theory to determine the confinement dimensionality of every band in a given system.
This scaling theory is valid for any type of physical wave - acoustic, electromagnetic, electron, spin, \textit{etc}. - in both quantum and classical setting, and for systems in any dimension, and is readily usable in computer algorithms, allowing for automated classification of the bands.

Nevertheless, the theory of Ref.~\cite{Kozon2022Phys.Rev.Lett.} requires for every investigated superlattice a smaller \emph{reference} superlattice, so that one can observe the scaling behavior of the key quantities when changing the supercell size.
Generally, obtaining the data for the reference supercell is significantly less computationally demanding than performing the calculations for the supercell of interest.
On the other hand, this requirement for a reference supercell makes the scaling approach of Ref.~\cite{Kozon2022Phys.Rev.Lett.} inapplicable in case the reference supercell is not available.
Furthermore, the scaling theory is exact only in the case of infinitely large supercells and thus for experimentally relevant small supercells inaccuracies may occur, as described in~Ref.~\cite{Kozon2022Phys.Rev.Lett.}.
Since the reference supercell is smaller than the investigated one, it is clear that eliminating the reference supercell from the confinement determination may provide an advantage in terms of accuracy of the scaling method.

Assigning the confinement dimensionality to each band in a spectrum is a typical classification task, which is among the archetypal problems solved by unsupervised machine learning~\cite{Dutton1997Knowl.Eng.Rev.}.
A prominent approach within unsupervised learning are clustering algorithms, which aim to partition the dataset into several clusters in such a way that ``similar'' data points are grouped together in the same cluster~\cite{Jain1988,Jain1999ACMComput.Surv.}.
In the context of wave characterization, machine learning has been employed, \eg , by Refs.~\cite{Leykam2021APLPhotonics,Scheurer2020Phys.Rev.Lett.} for topological classification of band structures pertaining to various types of periodic lattices.

In this paper, we investigate the application of clustering algorithms to improve the precision of confinement identification for waves in small supercells.
We employ two clustering algorithms to partition the bands of interest in the two-dimensional parameter space of mode volume and confinement energy, so that each resulting cluster corresponds to a specific confinement dimensionality.
The first one is the well-known k-means algorithm~\cite{Macqueen1967} with improved initialization~\cite{Arthur2007Proc.EighteenthAnnu.ACM-SIAMSymp.DiscreteAlgorithms}, which measures the ``similarity'' between the bands by their distance in a parameter space.
This is a standard algorithm, which, however, does not take into account any specific physical properties of our system and, as we show, this results in less accuracy in some cases.
Secondly, we implement our own model-based clustering algorithm (MBC), which uniquely merges clustering with the physical model behind the wave confinement, measuring the ``similarity'' between the bands by their distance from the scaling curves predicted by Ref.~\cite{Kozon2022Phys.Rev.Lett.}.
The MBC algorithm shows improved accuracy in the cases where k-means fails.

We conclude that even though the clustering approach brings specific complications on its own, it can be used to enhance the precision of confinement identification for small supercells, especially if one first employs the direct scaling method of Ref.~\cite{Kozon2022Phys.Rev.Lett.} to determine the correct set of confinement dimensionalities.
In terms of computational complexity, the additional clustering step is negligible, taking mere seconds for the cases presented in this paper.

The remainder of this paper is organized as follows:
Section 2 of this paper introduces notation and a brief review of scaling theory of wave confinement.
In the third section, we reformulate the scaling equation for wave confinement into a more machine-learning friendly form.
The fourth section deals with the evaluation of the accuracy of the clustering algorithms, while in the fifth section we describe our model-based clustering algorithm.
In the sixth section, we compare the performance of the k-means clustering algorithm, our MBC algorithm, and the standard scaling approach using a reference supercell.
 
\section{Superlattices and scaling}
\begin{figure}[htbp]
    \centering
    \includegraphics[width=\textwidth]{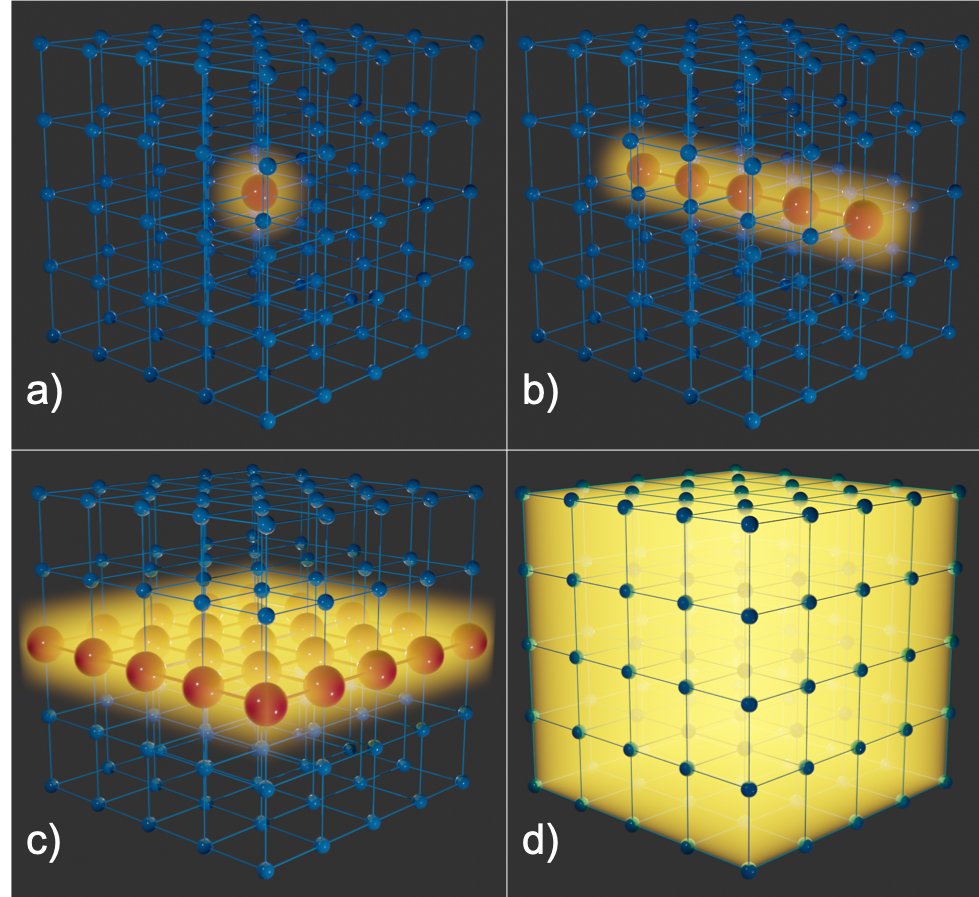}
    \caption{Illustration of 3D supercells containing various defects that contain waves. 
    Blue spheres represent unperturbed unit cells, red spheres correspond to defect unit cells and confined waves are depicted in yellow. 
    (a) Supercell containing a point defect, $c=3$. 
    (b) Supercell containing a line defect, $c=2$. 
    (c) Supercell containing a plane defect, $c=1$.
    (d) Supercell with no defects, where all unit cells are identical, does not support confined waves, $c=0$.}
    \label{fig:supercells}
\end{figure}

Superimposing a periodic lattice of defects on another underlying crystal lattice gives rise to a so-called superlattice~\cite{Bragg1934Proc.R.Soc.Lond.A,Bethe1935Proc.R.Soc.Lond.A,Ivchenko1997,Kozon2022Phys.Rev.Lett.}. 
A unit cell of such a superlattice is called a supercell. 
Fig.~\ref{fig:supercells} depicts supercells containing various types of defects. 
This is a traditional model for, \eg , defects in periodic materials. 
A supercell of linear size $N$ in a $D$-dimensional system contains $N^D$ unit cells. 
A defect with a dimensionality $d$ can confine waves in $c=D-d$ dimensions, as illustrated in Fig.~\ref{fig:supercells}. 
The number $c$ is referred to as confinement dimensionality.

In Ref.~\cite{Kozon2022Phys.Rev.Lett.}, we have developed a general scaling theory of wave confinement. 
This method relies on the confinement quantity $W(\vb r)$ as an input. 
This quantity depends on the investigated system and represents what is intuitively understood as confinement; it can be, \eg , the energy density in photonic systems, or the charge density in electronic systems. 
From $W(\vb r)$, two other quantities are calculated;
the relative confinement energy defined as 
\begin{equation}\label{confineenergy}
    \EE \coloneqq \frac{\int_{\Vcav} W(\vb r) \dd V}{\Esup},
\end{equation}
where the quantity $\Esup\coloneqq\int_\Vsup W(\vb r)\dd\vb r$ represents the total energy in the supercell,
and the relative mode volume 
\begin{equation}\label{modevolume}
    \VV = \Vmode \coloneqq \frac{\int_{\Vcav} W(\vb r) \dd V}{ \Vsup \max\limits_{\rr\in\Vsup} \{W  (\vb r)\}}.
\end{equation}
Here, $\Vsup$ is the supercell volume and $\Vcav$ a specific integration volume, usually chosen so that it contains a part of the defect.
The quantities defined by Eqs.~\eq{confineenergy} and~\eq{modevolume} have the following scaling behavior in the limit $N\rightarrow\infty$:
\begin{equation}
\VV=A N^{-c}, \qquad \EE=B N^{c-D},
\label{vescaling}
\end{equation}
with $A,B$ constants that are independent of $N$.

We found that by observing the scaling behavior of the combined quantity
\begin{equation}
\frac{\VV^\alpha}{\EE} = C N^\kappa
\label{veratio}
\end{equation}
with respect to the supercell size $N$, for judiciously chosen auxiliary powers $\alpha$, one can determine the confinement dimensionality $c$ for each band of states of a given system~\cite{Kozon2022Phys.Rev.Lett.}. 
Here, the constant $C=A^\alpha/B$ is \textit{a priori} unknown and may depend on various system parameters and the specific investigated band.
Although this scaling approach is rigorously valid only for sufficiently large $N$, it was found that it works for several smaller systems as well, which is unusual for scaling theories.
However, the method sometimes also returned unphysical results for certain bands in these small supercells, necessitating a critical evaluation of the results.

In this paper, we present an alternative scaling approach and employ machine learning algorithms to enhance the precision of the method of Ref.~\cite{Kozon2022Phys.Rev.Lett.}.
Although this approach introduces an additional step in the confinement analysis, this step is computationally cheap and improves the accuracy of the results for small supercells. 
Moreover the clustering analysis still maintains its wide range of applicability to different physical systems.

Throughout this paper, we use as reference system a 3D inverse woodpile photonic band gap crystal with two proximate defects \cite{Ho1994SolidStateCommun.}, described in detail in Appendix~\ref{iwpc}.
Specifically, we employ the well-researched structure with regular pores of radius $R=0.24a$, with $a$ being the lattice parameter, and defect pores of radius $R^\pr=0.5R$~\cite{Woldering2014Phys.Rev.B, Hack2019Phys.Rev.B, Devashish2019Phys.Rev.B}.
This system represents a suitable model due to its relatively complex arrangement of linear defects, creating a cavity at their crossover.
We know that, as a result, the system exhibits extended ($c=0$), line-confined ($c=2$), and point-confined ($c=3$) bands that need to be distinguished, corresponding to the total number of $K=3$ clusters.
On the other hand, we know from physics that the system does not support plane-confined ($c=1$) bands and those should therefore not appear in the correct clustering result.
We employ the range of supercell sizes $N=2,3,4$, exhausting what is feasible in a reasonable computing time.
For every supercell size $N$, we used the plane-wave expansion method implemented via the well-known MPB code~\cite{Johnson2001Opt.Express} to obtain the mode volume and energy density for each band. 

\section{Machine-learning approach to scaling}
\label{sec:MLdef}

\begin{figure}[htbp]
    \centering
    \includegraphics[width=\textwidth]{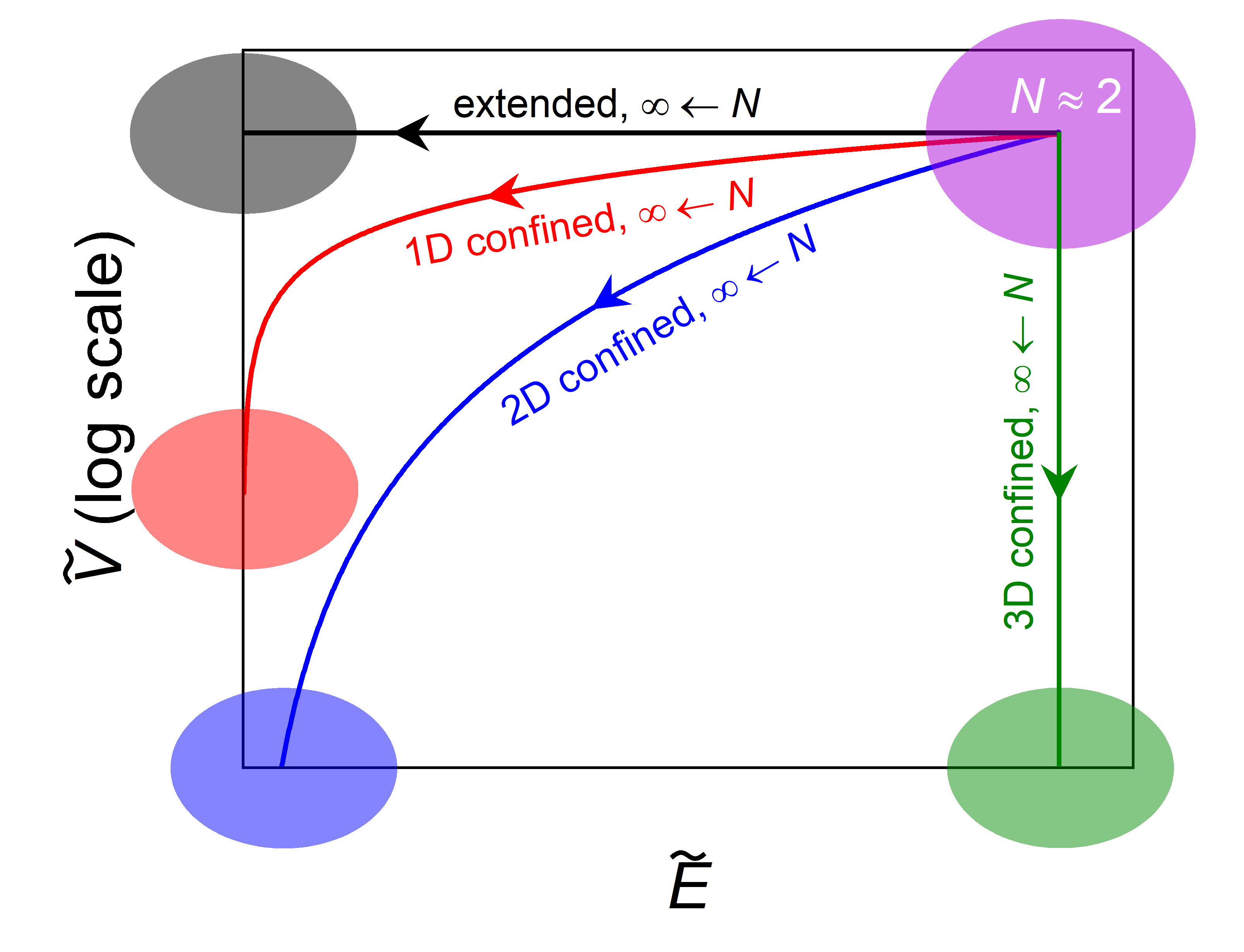}
     \caption{Schematic of the scaling behavior of the bands according to Eq.~\eq{MVEC}. 
     Mode volume versus the confinement energy. 
     For small supercell size $N$ all the bands are grouped together, but, for larger N, they spread over the $(\EE,\VV)$ space, forming clusters in accordance with their confinement dimensionality $c$.}
    \label{fig: scaling}
\end{figure}

Instead of combining the quantities in Eqs.~\eq{confineenergy} and~\eq{modevolume} into the ratio~\eq{veratio}, one can also express the mode volume as a function of confinement energy, in the limit $N\rightarrow\infty$:
\begin{equation}
\VV = \widetilde C \EE ^{\frac{-c}{c-D}},
\label{MVEC}
\end{equation}
where
\begin{equation}
\widetilde C = \frac{A}{B^{\frac{-c}{c-D}}}.
\end{equation}
Eq.~\eq{MVEC} represents a different form of scaling than Eq.~\eq{veratio}, this time with respect to $\EE$.
Of course, from Eq.~\eq{vescaling}, it is clear that $\EE$ also changes with the system size $N$.
The band behavior based on Eq.~\eq{MVEC} is graphically illustrated in Fig.~\ref{fig: scaling}.
For small supercell sizes $N$, the bands are accumulated in the upper right corner with relatively high $\EE$ and $\VV$.
As the supercell size increases, the bands start to follow a path corresponding to their confinement dimensionality $c$.
Eventually, bands with the same $c$ form clusters as depicted in Fig.~\ref{fig: scaling}.
The bands in a $D$ dimensional system will form at most $D + 1$ clusters, corresponding to all possible values of the confinement dimensionality $c$.

The immediate advantage of this approach compared to the one presented in Ref.~\cite{Kozon2022Phys.Rev.Lett.} is that this method does not require a smaller reference supercell, but one can directly analyze which bands belong to which cluster for the supercell size of interest.
Nevertheless, since Eq.~\eq{MVEC} strictly holds only for $N \to \infty$, the clusters will be clearly distinguishable only for sufficiently large supercells.
In both computations and experiments, one is realistically constrained to relatively low $N$ and it thus may become difficult to distinguish the clusters from each other.
To improve the accuracy of confinement determination, we propose here the employment of a clustering algorithm, which divides the data into clusters quantitatively and automatically.

\subsection{Data clustering}
Data clustering is one of the staple problems of machine learning techniques~\cite{Dutton1997Knowl.Eng.Rev.} and it is thus reasonable to expect that such techniques enhance the precision of confinement identification.
In order to formulate the band confinement identification as a clustering problem, we consider the logarithm of mode volumes $\log_{10}\VV_i$ and the confinement energies $\EE_i$ for each band $i=1,\ldots, M_N$, where $M_N$ is the total number of bands for the corresponding supercell. 
To ensure that both $\log_{10}\VV_i$ and $\EE_i$ are treated with equal importance and thus neither dominates the clustering, we renormalize the data once again, so that both variables have the same range of values prior to clustering.
The normalization is performed for each dataset separately as follows:
\begin{equation}\label{ecrenorm}
    \mathcal E_i= \frac{\EE_i}{\max\limits_{1\le j \le M_N} \left\{\EE_j\right\}-\min\limits_{1\le j \le M_N} \left\{\EE_j\right\}},
\end{equation}
\begin{equation}\label{mvrenorm}
    \mathcal V_i= \frac{\log_{10}\VV_i}{\max\limits_{1\le j \le M_N} \left\{\log_{10}\VV_j\right\}-\min\limits_{1\le j \le M_N} \left\{\log_{10}\VV_j\right\}}.
\end{equation}
Note that we choose the semi-logarithmic variable space because this provides more uniformly spread data than both linear and log-log view.
The normalized dataset used for the clustering is thus $\XX_N=\{\vb x_i \eqdef (\mathcal{E}_i,\mathcal{V}_i)\,|\,i=1,\ldots,M_N\}$.

The goal of a clustering algorithm is to subdivide the data in $\XX_N$ among $K$ clusters, such that ``similar'' data points are assigned to the same cluster $C_k=\{\vb x^k_j\,|\,j=1,\ldots, M_k\}$, where $k=1,\ldots,K$.
In total, each cluster will contain $M_k\ge 1$ data points, such that $\sum_{k=1}^K M_k=M_N$.
We refer to the specific distribution of the data points within the clusters as a \emph{partition}, denoting it as $\CCC_K=\{C_1,\ldots,C_K\}$.
The centroid $\vb c_{k}$ of the cluster $C_k$ is defined as the mean of its constituent points, \ie , $\vb c_{k}=(\sum_{j=1}^{M_k} \vb x_j^k)/M_k$.
We denote the set of all possible partitions of the dataset $\XX_N$ into $K$ clusters as $\PPP_K$.

We explore two different ways of clustering the data for the purpose of confinement identification:
A standard k-means++ algorithm implemented in the Python library \emph{scikit-learn}~\cite{Pedregosa2011J.Mach.Learn.Res.} and our own MBC algorithm utilizing physical insight to perform the clustering.
Each of them utilizes a slightly different notion of what it means for two data points to be ``similar'', and therefore employs a different approach to obtain the resulting partition, as described below.

\subsection{The k-means++ algorithm}

A well-known known clustering algorithm is the k-means algorithm~\cite{Macqueen1967}, which considers two data points to be ``similar'' if their Euclidean distance in the data space is small.
Therefore, it aims to group such points together in one cluster.
Within a given partition $\CCC_K$, this notion of similarity can be measured via the cost function $\Psi:\PPP_K\ra\RR$ representing the sum of distances of each data point from the centroid of its cluster:
\begin{equation}\label{Oobj_func}
\Psi(\CCC_K) \coloneqq \sum^{K}_{k=1} \sum_{\vb x\in C_k} \norm{\vb x - \vb c_{k}}^2,
\end{equation}
where $\norm{\cdot}$ denotes the Euclidean norm.
The k-means algorithm aims to minimize the cost function $\Psi$ over all possible partitions.
The partition $\tilde\CCC_K$, for which $\Psi$ attains its minimum is considered the best partition of the dataset, \ie , $\tilde\CCC_K$ is such that
\begin{equation}\label{Oobj_funcmin}
\Psi(\tilde\CCC_K) = \min_{\CCC_K\in\PPP_K} \Psi(\CCC_K).
\end{equation}
The k-means algorithm is summarized in Algorithm~\ref{kmeans}.
\begin{algorithm}
\caption{k-means algorithm}\label{kmeans}
\begin{description}[font=\bfseries, align=left, leftmargin=1.45cm]
\item[Input:] Data array $\XX_N \subset \RR^2$,  number of clusters $K \in \NN$.
\item[Output:] Division of the data into $K$ clusters such that the cost function $\Psi$ is minimized.
\end{description}
\begin{enumerate}[font=\bfseries, align=left]
\item Arbitrarily select the initial centroids $\vb c_k$ for each $k= 1,\ldots, K$.
\item Assign each data point $\vb x_i \in \XX_N$ to its nearest centroid using the Euclidean metric. 
In case of equality of the nearest centroids, the data point is assigned to either cluster.
\item Recalculate the centroid of each cluster as the mean of all its constituent data points.
\item Repeat the steps 2 and 3 with the new centroids until a termination criterion is reached.
\end{enumerate}
\end{algorithm}

Algorithm~\ref{kmeans} will always converge to a local minimum of the cost function $\Psi$, see Ref.~\cite{Piccialli2022INFORMSJ.Comput.}. 
There is, however, a likelihood that this local minimum does not coincide with the global minimum. 
As a result, the output of the algorithm may depend on the specific initialization choice.

One can improve the probability of reaching the global minimum by applying the algorithm multiple times with different random initializations and choosing the result with the smallest cost function $\Psi$.
Alternatively, one can choose a more sophisticated method of the cluster centroid initialization, as described by the k-means++ algorithm~\cite{Arthur2007Proc.EighteenthAnnu.ACM-SIAMSymp.DiscreteAlgorithms}.
Here, instead of choosing the initial centroids in Step 1 of Algorithm~\ref{kmeans} at random, we choose the first centroid randomly from the dataset $\XX_N$, with a uniform probability distribution.
Then, we continue to choose each subsequent centroid $\vb c_k, 1<k\le K$ randomly from the dataset $\XX_N$ with the adjusted probability distribution
\begin{equation}
P_k(\vb x)=\frac{D_k^2(\vb x)}{\sum_{\vb x^\pr\in\XX_N}D_k^2(\vb x^\pr)},
\end{equation}
where 
\begin{equation}
D_k(\vb x)=\min_{1\le j < k} \left\{\norm{\vb x - \vb c_j}\right\}
\end{equation}
is the distance of the point $\vb x$ to the closest centroid that has already been determined.
The remaining part of the k-means++ algorithm is the same as for the standard k-means algorithm, described in Algorithm~\ref{kmeans}.

\section{Clustering accuracy}
Clustering of the bands based on their confinement dimensionality is complicated by the fact that there is no known ground truth, \ie , there is no reference solution against which we can assess the accuracy of the clustering process and evaluate its validity.
This is an important hurdle for the clustering approach, since the k-means++ algorithm requires the number of clusters $K$ as an input, which is, however, not known \textit{a priori}.
One can, of course, run the algorithms for all possible values of $K$, but how can we determine that we have found the correct number of clusters?
Since the cost function $\Psi$ defined by Eq.~\eq{Oobj_func} is, by definition, a decreasing function of cluster numbers, it cannot be used to determine the correct $K$.
We therefore need another type of clustering accuracy measure and this is where the concept of cluster validity indices (CVIs) comes into play.

Cluster validity indices are quantitative measures to determine the validity of the clustering results.
A plethora of CVIs has been proposed in literature, employing various approaches to measuring the clustering accuracy~\cite{Arbelaitz2013PatternRecognition}. 
Based on the definition of each CVI, better clustering results correspond to either lower or higher values of the index and thus the most accurate clustering out of a set of results is the one achieving the optimum of the CVI, \ie , either the minimum or the maximum.
CVIs can be divided into two categories in a rather straightforward manner: external indices, comparing the clustering result against the ground truth, and internal indices, analyzing only the partitioned data without the need for any external information~\cite{Saxena2017Neurocomputing}.
It is clear that for our purposes we are interested in the CVIs of the internal type.
Most CVIs measure \emph{cluster cohesion} and \emph{cluster separation} of the data, in some sense. 
Cluster cohesion describes to what extent the entities inside a cluster are alike, whereas cluster separation evaluates how well different clusters are separated.
The behaviour of the CVIs largely depends on the context and the setting~\cite{Arbelaitz2013PatternRecognition}.
This is why numerical tests must be conducted to convincingly select the CVI that provides the correct measure of accuracy in the given context~\cite{Arbelaitz2013PatternRecognition}. 
This obviously also applies to clustering for wave-confinement analysis.
Therefore, in this section, we analyze select CVIs, in order to find the ultimate clustering accuracy measure for our specific problem.




\subsection{Methodology}

To limit the number of CVIs to be tested, we pick five of the best-performing CVIs from Ref.~\cite{Arbelaitz2013PatternRecognition}. 
These are the Silhouette coefficient (Sil), the Calinski-Harabasz (CH), the Davies-Bouldin (DB) including its slight variation (DB*), the COP and the S\_Dbw indices.
Sil, CH, and DB were implemented via the package library \emph{scikit-learn}, S\_Dbw was implemented via the package library \emph{S\_Dbw}, and DB* and COP were implemented directly by ourselves.
For the overview of the CVIs and their definitions, see Appendix~\ref{CVIs}.

The testing procedure was executed in an identical fashion for each CVI, utilizing the reference inverse woodpile structure with $R=0.24a$ and $R^\pr=0.5R$, specified in Section 2 and Appendix~\ref{iwpc}.
To find the CVI that provides the best accuracy measure, we run the clustering algorithm for various total number of clusters $K$ and subsequently compute the value of each CVI for that partition.
The global optimum (minimum or maximum, depending on the CVI definition) should then be attained for the correct number of clusters $K=3$.
For this analysis, we choose the range of $K=2,\ldots,25$.
This range exceeds the number of clusters expected in the context of wave confinement, but such a broad range of values may provide additional information on the accuracy of the CVIs. 
Note that we do not use $K=1$ since the value of the CVIs is not defined for only one cluster, see their definitions in Appendix~\ref{CVIs}. 
It is important to note that the choice of clustering algorithm should not affect the CVI performance, since CVIs only evaluate the intrinsic clustering outcome~\cite{Rousseeuw1987JournalofComputationalandAppliedMathematics, Davies1979IEEETrans.PatternAnal.Mach.Intell.}.
This is why, for the purpose of the CVI analysis, we stick to only the k-means++ algorithm.

According to the supercell scaling method the distance between the clusters of different levels of confinement should increase as the supercell size $N$ grows. 
Taking also into account that we always normalize the dataset to a unit domain, we expect the well-behaved CVIs to display better result, \ie , lower minimum or higher maximum, every time $N$ is incremented. 
In other words, the CVI must be a monotonically increasing or decreasing function of $N$, depending on the CVI definition. 
We refer to this as the convergence property of the CVI. 

Thus, in this section, we are looking for the best performing CVI for our specific setting of wave confinement analysis. 
The suitable CVI has to satisfy the following criteria:
\begin{enumerate}
    \item For the reference structure, the maximum or minimum of the CVI should be obtained at $K=3$ clusters.
    \item The CVI should be a monotonically increasing or decreasing function of the supercell size $N$, exhibiting convergence.
\end{enumerate}

\subsection{Results and discussion}
\begin{figure}[htbp]
\centering
    \includegraphics[width=0.97\textwidth]{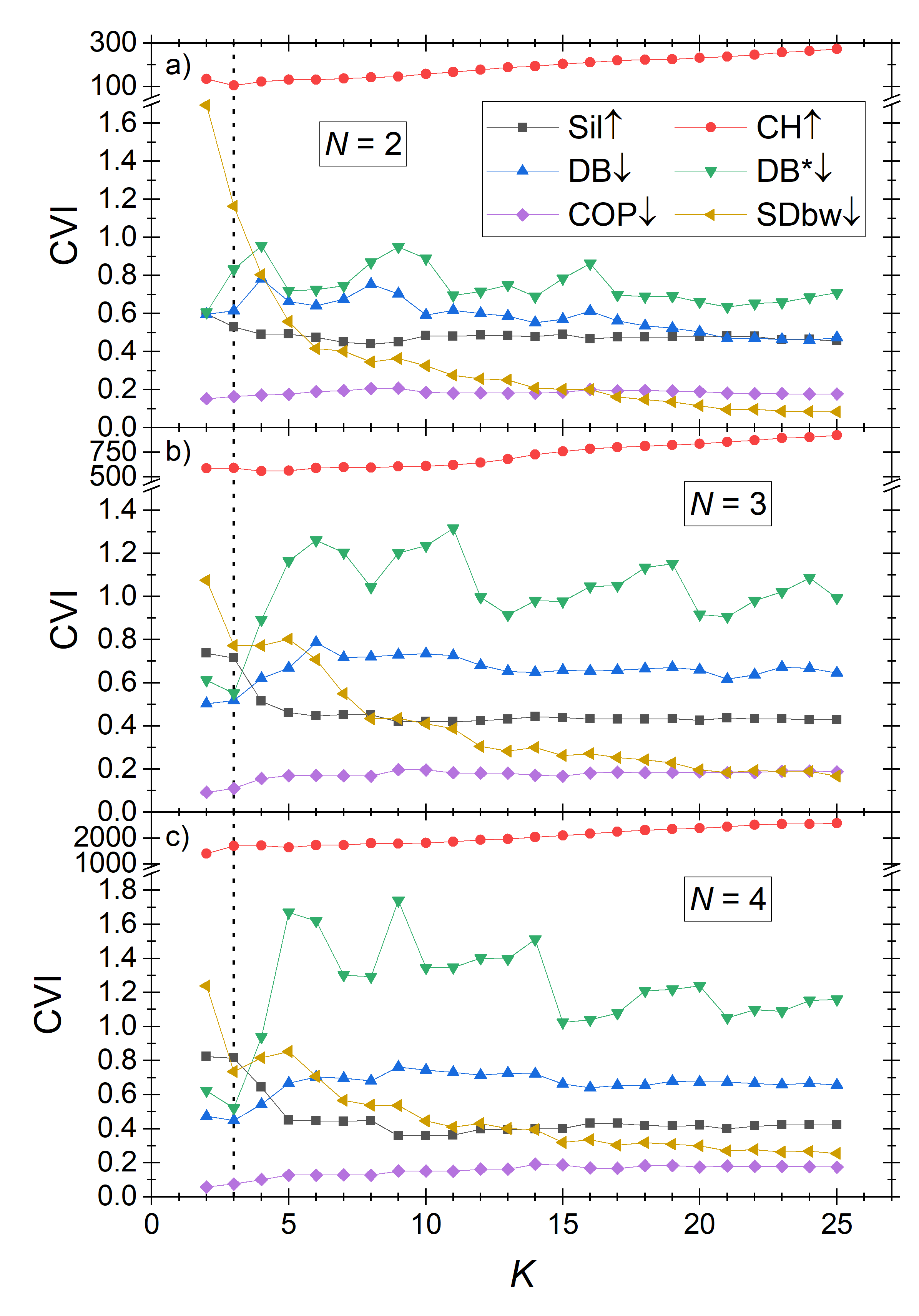}
  \caption{CVI values as a function of the number of clusters $K$. The arrows in the legend indicate if the optimum of a specific CVI is a minimum (down arrow) or a maximum (up arrow). The suitable CVI should attain its respective optimum at the correct number of clusters $K=3$, highlighted with the dotted vertical line. Note that for $K=1$ the CVIs are undefined. (a) $N=2$ supercell. (b) $N=3$ supercell. (c) $N=4$ supercell.}
  \label{fig:CVI_K}
\end{figure}

Fig.~\ref{fig:CVI_K} shows obtained CVI values as a function of the input number of cluster $K$, for the reference structure.
For the supercell size $N=2$, none of the indices achieve their respective optimum for $K=3$.
This is not surprising, as this is an extremely small supercell and the scaling properties are not yet well developed for this size.
In the case of $N=3$, the only CVI achieving correctly its optimum is DB* at $K=3$.
Finally, for $N=4$, both DB and DB* achieve their minimum for the correct value $K=3$.
DB and DB* also exhibit very similar behavior with respect to the number of clusters $K$ and certain qualitative difference between them is only visible in the regime of large $K$, where DB stays mostly constant around a value relatively close to its minimum, whereas DB* oscillates around a value much greater than its minimum.
The other CVIs exhibit rather erratic behaviour. 
CH  and S\_Dbw  exhibit local, but not global optima at $K=3$, whereas Sil and COP do not even exhibit local optima at the correct number of clusters.

\begin{figure}[htbp]
\centering
    \includegraphics[width=\textwidth]{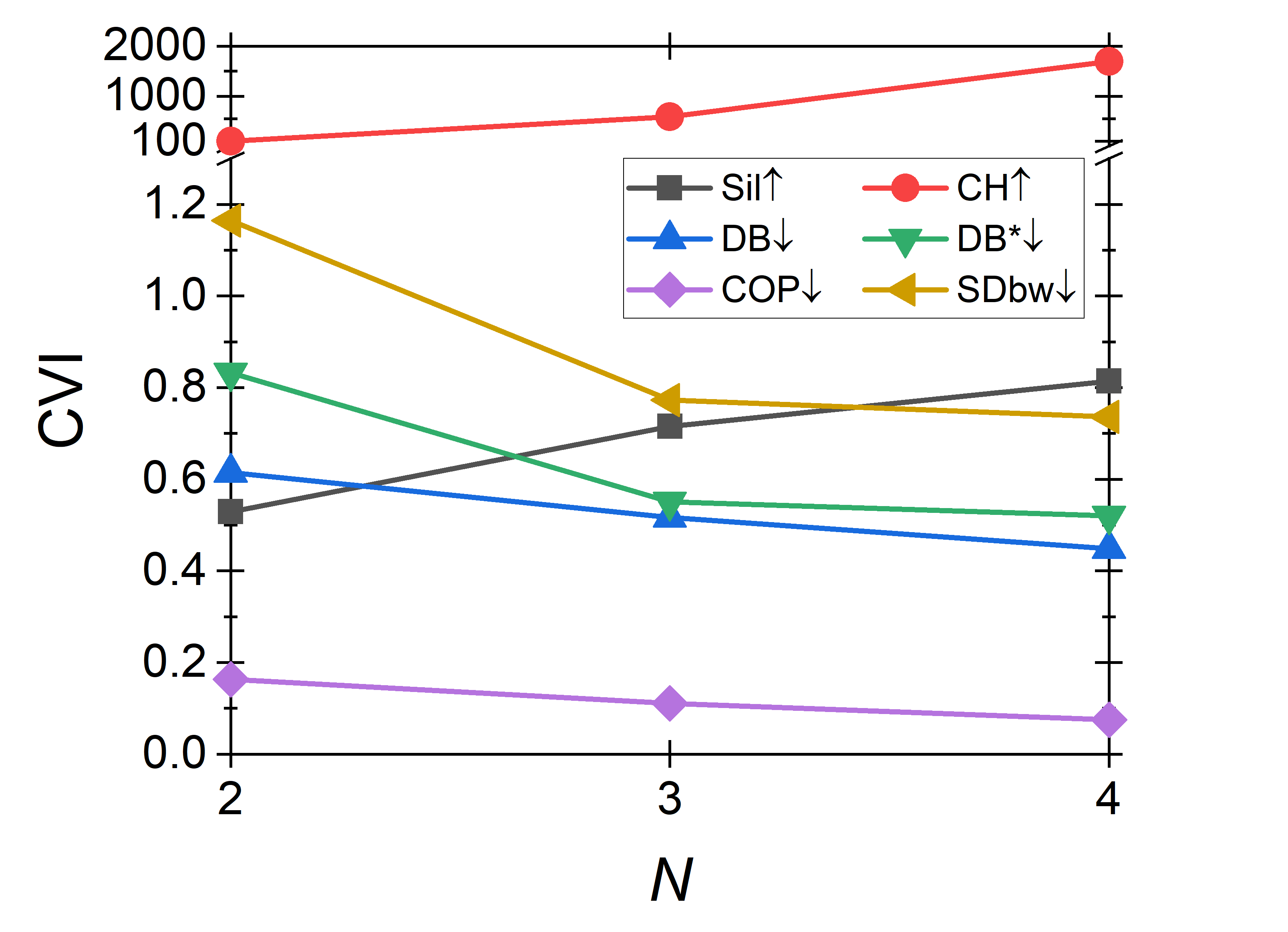}
  \caption{CVI values as a function of the supercell size $N$ for the correct number of clusters $K=3$. The arrows in the legend indicate if the optimum of a specific CVI is a minimum (down arrow) or a maximum (up arrow). The suitable CVI is expected to decrease (increase) with growing $N$, if its optimum is a minimum (maximum).}
  \label{fig:CVI_N}
\end{figure}

Fig.~\ref{fig:CVI_N} shows the convergence behavior of CVIs with respect to the supercell size $N$. 
Here, it is apparent that all the indices obey the required monotonic convergence.

Based on our analysis, we conclude that, out of our tested sample of CVIs, DB* is the only CVI to satisfy the requirements for a good clustering accuracy measure in our setting for both $N=3$ and $N=4$ supercell sizes.
Therefore, we will focus on this CVI in the remainder of this paper.
We also note that for the larger supercell size $N=4$, DB also seems to be well-behaved and thus might be suitable as a CVI for larger supercell sizes.

\section{Model-based clustering}

The k-means clustering algorithm is a standard and easy-to-implement process.
However, it simply clusters the data without any account for the underlying physics.
Therefore, we present here our model-based clustering (MBC) algorithm, a unique mix of clustering and a model-based regression method.

As discussed in Section~\ref{sec:MLdef}, bands in a superlattice analysis follow certain trajectories in the $(\EE,\log_{10}\VV)$ space based on their confinement dimensionalities $c$, forming clusters as depicted in Fig.~\ref{fig: scaling}.
Mathematically, upon transforming Eq.~\eq{MVEC} from the $(\EE,\VV)$ to $(\EE,\log_{10}\VV)$ space, the trajectory of a band with confinement dimensionality $c$ will be given by 
\begin{equation}
   \log_{10} \VV = \log_{10}\tilde C + \frac{c}{D-c} \log_{10}\EE.
   \label{MBCcurve0}
\end{equation}
Note that the second normalization given by Eqs.~\eq{ecrenorm} and~\eq{mvrenorm} only adds a constant to the right-hand side of Eq.~\eq{MBCcurve0}, which can be absorbed by the unknown constant $\tilde C$.
We can thus directly write
\begin{equation}
 \mathcal V = \log_{10} \tilde C + \frac{c}{D-c} \log_{10}\mathcal E.
   \label{MBCcurve}
\end{equation}

In the MBC method, instead of evaluating the distance of a data point to the centroid of its cluster, we evaluate the distance of a point to the curve given by Eq.~\eq{MBCcurve}.
For $0\le c\le D$, we define the distance of a point $\vb x_0=(\mathcal E_0,\mathcal V_0)$ to a point $\vb x=(\mathcal E,\mathcal V)$ on the curve~\eq{MBCcurve} as
\begin{equation}
   \delta_c(\vb x_0) \coloneqq \min_{\vb x\in\gamma_c} \norm{\vb x-\vb x_0},
   \label{MBCdistance}
\end{equation}
where
\begin{equation}
       \gamma_{c} \coloneqq \{(\mathcal E, \mathcal V) \text{, satisfying \eq{MBCcurve}}\}.
   \label{gammacurve}
\end{equation}
Note that $\gamma_{c}$ formalizes the curves in Fig.~\ref{fig: scaling}: In a $D=3$ system, $\gamma_0$ corresponds to the black, $\gamma_1$ to the red, $\gamma_2$ to the blue and $\gamma_3$ to the green curve, for specific choices of the scaling constants $\tilde C$.
For $0\le c < D$, the norm on the right hand side of Eq.~\eq{MBCdistance} can be explicitly written as
\begin{equation}
   \norm{\vb x-\vb x_0} = \sqrt{(\mathcal E- \mathcal E_0)^2+ ( \log_{10} \tilde C + \frac{-c}{c-D} \log_{10} \mathcal E - \mathcal V_0)^2},
   \label{MBCnorm}
\end{equation}
and, for $c=D$, it is simply given by the difference in the mode volumes $\abs{\mathcal V-\mathcal V_0}$.

In the MBC algorithm, we calculate the distance of each data point to each curve corresponding to different values of $c$ and assign the data point to the cluster corresponding to the smallest distance.
Based on this clustering, we then calculate the centroids of each cluster and adjust the curves $\gamma_c$ to pass through these new centroids by changing $\tilde C$ in \eq{MBCcurve}.
Then the algorithm is iterated again, until a termination criterion is reached.
Since, for sufficiently large supercells, curves $\gamma_c$ are strictly separated from each other, \ie , they do not cross, as illustrated in Fig.~\ref{fig: scaling}, there is no need for repeated random initialization.
As initialization values, we simply set all the centroids at the point $(\max_{i=1}^{M_N}\mathcal E,\max_{i=1}^{M_N}\mathcal V)$, corresponding to the top-right corner of the clustering dataset in Fig.~\ref{fig: scaling}.

The MBC algorithm is summarized in Algorithm~\ref{MBCalg}.
\begin{algorithm}
\caption{Model-based clustering algorithm}\label{MBCalg}
\begin{description}[font=\bfseries, align=left, leftmargin=1.15cm]
\item[Input:] Data array $\XX_N \subset \RR^{2}$,  values of $K$ confinement dimensionalities, corresponding to different clusters.
\end{description}
\begin{description}[font=\bfseries, align=left, leftmargin=1.45cm]
\item[Output:] Division of the data into clusters such that the distance from each point to its corresponding curve is minimized.
\end{description}
\begin{enumerate}[font=\bfseries, align=left]
\item Set the initial centroids to (1,0) for each curve.
\item Assign each data point $\vb x_i = (\mathcal{E}_i,\mathcal{V}_i) \in \XX_N$ to its nearest curve $\gamma_c$ using the distance measure $\delta_c$. 
In case of equality of the nearest centroids, the data point is arbitrarily assigned to one of the corresponding clusters.
\item Recalculate the centroid of each cluster as the mean of all its constituent data points and adjust the corresponding curve to pass through the new centroid.
\item Repeat the steps 2 and 3 with the new curves until a termination criterion is reached.
\end{enumerate}
\end{algorithm}
We note that, while the k-means algorithm only performs the clustering and one needs to manually assign the corresponding confinement dimensionality $c$ to each cluster, MBC does this inherently and automatically, which is an additional advantage of this approach.

Again, we use DB* as a CVI to determine the correct number of clusters.
We stress here that for MBC, the input is not only the number of clusters, but also their specific confinement dimensionalities.
The set of $K$ clusters can thus include different combination of $c$ values, yielding different partitions.
In such a case, DB* can sometimes prefer a physically impossible partition, such as plane-confined bands in a structure with no plane defects.
Nevertheless, if we restrict ourselves to only sets of physically meaningful confinement dimensionalities, the performance of DB* is comparable to the case of the k-means algorithm.
This is inherently a shortcoming of DB* applied to our problem, as it does not measure the validity with respect to the theory, but only based on clustering cohesion and separation, as discussed in more detail in Appendix~\ref{CVIs}.
There would thus be a clear benefit in devising a model-based CVI along with our model-based clustering algorithm.
This is, however, beyond the scope of this paper.
Moreover, as we discuss below, our computations suggest that the best accuracy of the confinement identification can be achieved by finding the correct set of confinement dimensionalities via the scaling method of Ref.~\cite{Kozon2022Phys.Rev.Lett.} combined with physical insight and use that as an input for the clustering algorithm. 
The use of a CVI can thus be skipped if one is able to perform the scaling and has some information about the physics of the investigated structure.

\section{Performance analysis in confinement classification}

In this section, we compare the results obtained by our MBC algorithm and by the k-means++ algorithm with the results obtained by direct application of the scaling method described in Ref.~\cite{Kozon2022Phys.Rev.Lett.} with reference supercell size $N_0=2$.
We do this by applying the methods to several inverse woodpile structures with different properties.
The absence of the ground truth information for these problems complicates the analysis and therefore we attempt to somewhat alleviate this issue by visually inspecting the energy-density distribution $W(\rr)$ of selected bands.
Note, however, that this visual inspection is a qualitative tool and it may be difficult to precisely assign the confinement dimensionality of certain bands, this way.

\begin{figure}[htbp]
\centering
    \includegraphics[width=\textwidth]{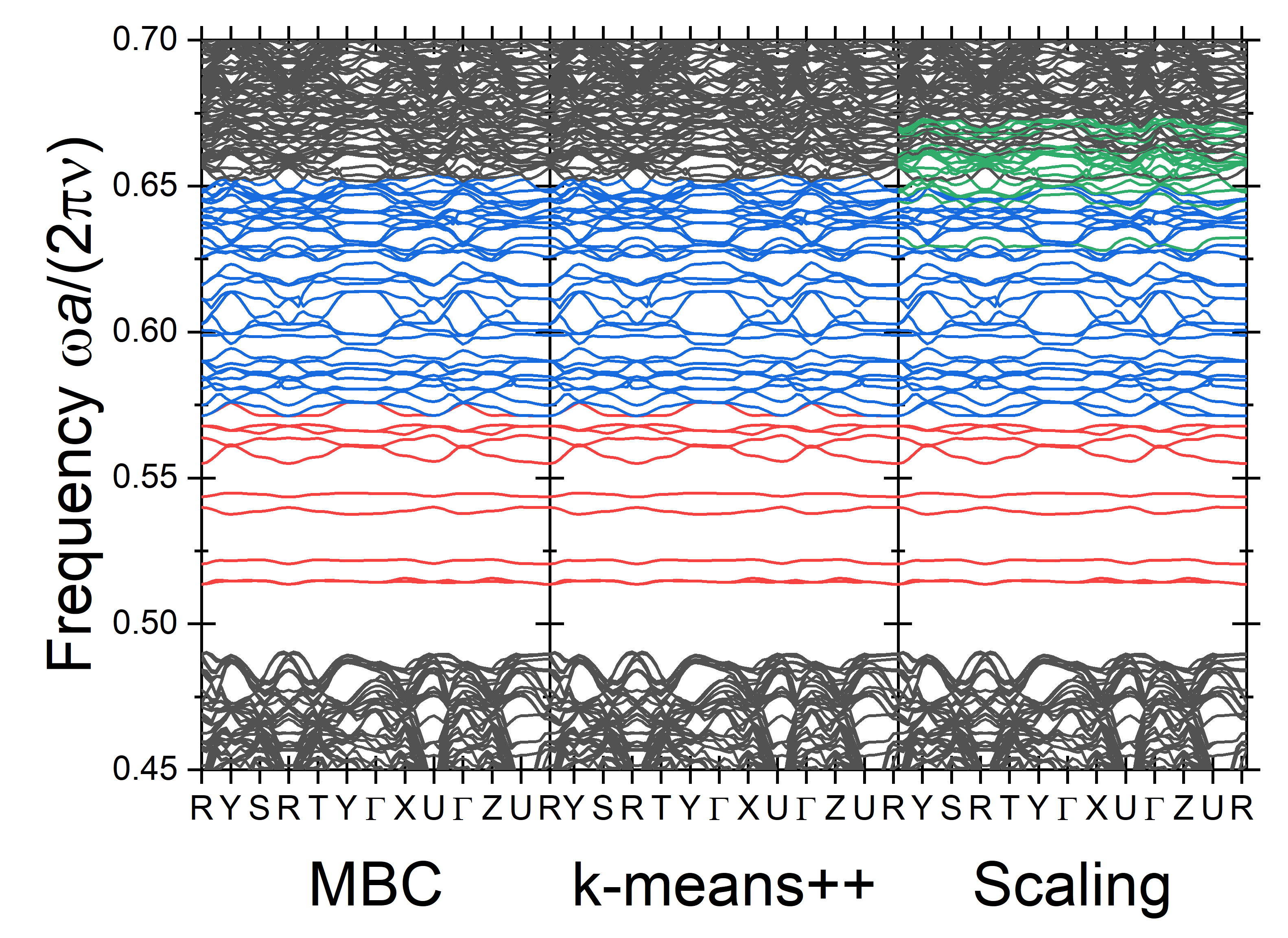}
  \caption{Band structure of an $N=3$ supercell of an inverse woodpile photonic crystal (see Appendix~\ref{iwpc}) with parameters $R=0.24a$, $R^\pr=0.5R$, with bands color-coded to indicate the confinement dimensionalities $c$ of each band. Red color corresponds to point-confined $c=3$, blue to line-confined $c=2$, green to plane-confined $c=1$ and black to extended $c=0$. The vertical axis is given in the reduced frequency $\tilde\omega=\omega a/(2\pi\nu)$, with $a$ being the lattice constant and $\nu$ the speed of light. The left panel shows the confinement classification for the MBC clustering, the central panel for the k-means++ clustering, and the right panel for the direct application of the scaling approach from Ref.~\cite{Kozon2022Phys.Rev.Lett.}.}
  \label{fig:N3r24rp05}
\end{figure}

Fig.~\ref{fig:N3r24rp05} shows the results of the classification of confinement dimensionality on an $N=3$ supercell of a silicon inverse woodpile photonic crystal with the most widely studied parameter set, namely an unperturbed pore radius $R=0.24a$, and a defect pore radius $R^\pr=0.5R$, see Appendix~\ref{iwpc} and Refs.~\cite{Woldering2014Phys.Rev.B, Hack2019Phys.Rev.B, Devashish2019Phys.Rev.B}. 
In this case, both clustering algorithms yield the same solution, which almost perfectly coincides with the direct scaling approach of~\cite{Kozon2022Phys.Rev.Lett.}. 
It is remarkable that all three methods independently agree on the presence of at least 9 bands with $c=3$, between $\tilde\omega\approx 0.52$ and $\tilde\omega\approx 0.57$, instead of the 5 previously identified by qualitative guesswork by Refs.~\cite{Woldering2014Phys.Rev.B,Devashish2019Phys.Rev.B,Hack2019Phys.Rev.B}.
Some of these bands have been shown to demonstrate ``Cartesian light'' or a 3D coupled cavity behavior\cite{Hack2019Phys.Rev.B} as also seen in recent experiments~\cite{AdhikaryInpreparation}.
The direct scaling approach of Ref.~\cite{Kozon2022Phys.Rev.Lett.}, however, identified several bands, mostly near $\tilde\omega\approx 0.65$, as $c=1$ plane-confined modes, which is unphysical, since the structure does not contain any planar defects, but only line and point ones.
Thanks to the fact that the clustering algorithms do not require a smaller reference supercell but work only with the larger investigated supercell, the clustering approach successfully avoids these unphysical results.
In this case, we therefore conclude that the clustering approaches are superior to the direct scaling.
It is important to stress that the scaling theory beyond the approach of Ref.~\cite{Kozon2022Phys.Rev.Lett.} still remains valid and the clustering merely improves its accuracy for small supercells.

\begin{figure}[htbp]
\centering
    \includegraphics[width=\textwidth]{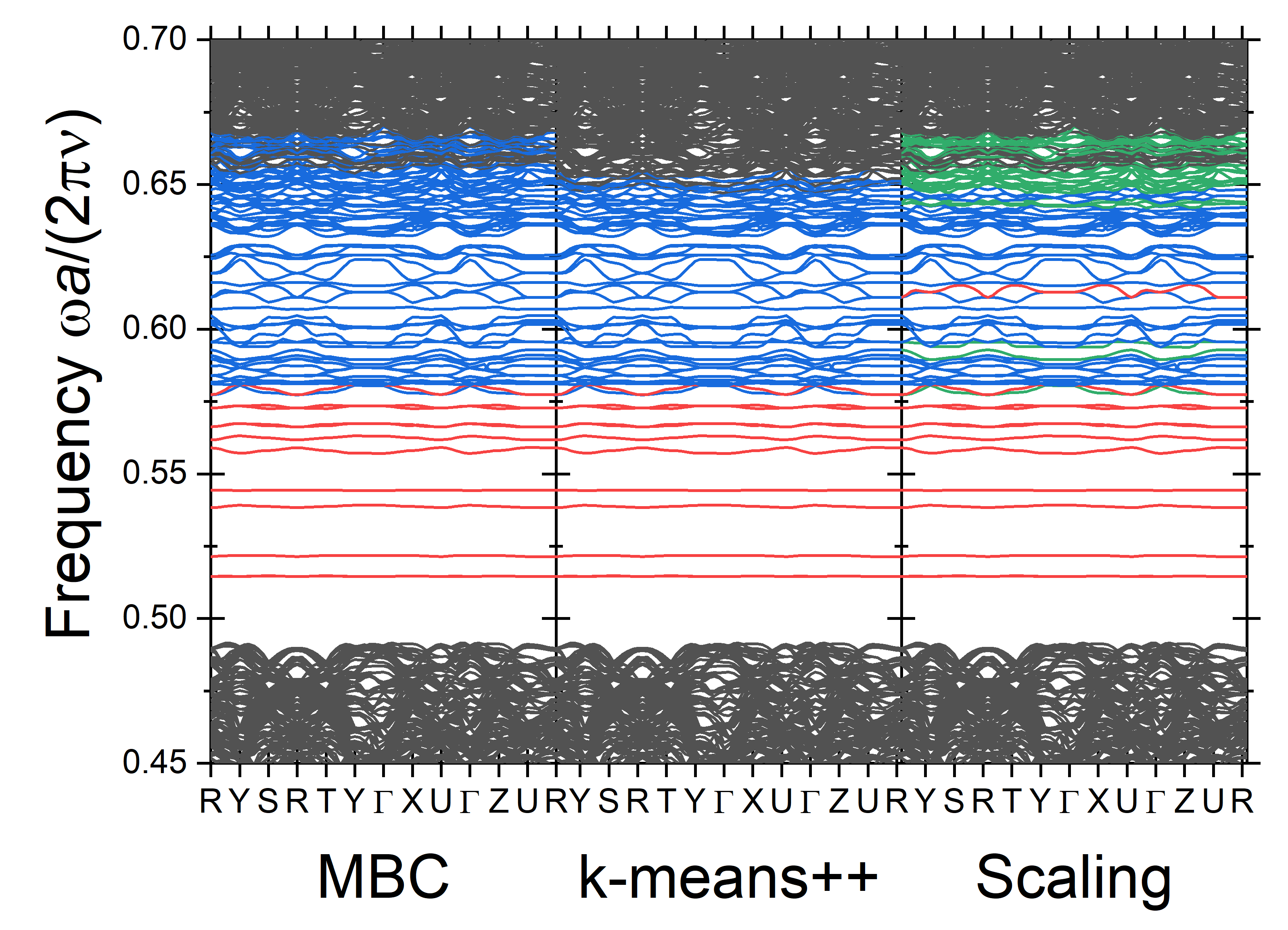}
  \caption{Band structure of an $N=4$ supercell of an inverse woodpile photonic crystal (see Appendix~\ref{iwpc}) with parameters $R=0.24a$, $R^\pr=0.5R$, with bands color-coded to indicate the confinement dimensionalities $c$ of each band. Red color corresponds to point-confined $c=3$, blue to line-confined $c=2$, green to plane-confined $c=1$ and black to extended $c=0$. The left panel shows the confinement classification for the MBC clustering, the central panel for the k-means++ clustering, and the right panel for the direct application of the scaling approach from Ref.~\cite{Kozon2022Phys.Rev.Lett.}.}
  \label{fig:N4r24rp05}
\end{figure}

Fig.~\ref{fig:N4r24rp05} shows the results of our confinement identification on a larger $N=4$ supercell of the inverse woodpile photonic crystal with the same unperturbed pore size $R=0.24a$, and defect pore size $R^\pr=0.5R$ as in the previous paragraph.
Note that all approaches agree that a higher number of $c=3$ bands is present here than for the $N=3$ supercell with the same pore sizes.
This clearly illustrates the problem with small supercells: the localization length of some confined bands is too large to be identifiable in the small supercells and hence these confined bands are only apparent with larger supercells.
In this case, the clustering approach again avoids the unphysical plane-confined bands ($c=1$) obtained by the direct scaling application near $\tilde\omega\approx 0.65$.
The k-means++ algorithm classifies the $c=2$ and the $c=3$ bands in a very similar way to the direct scaling approach.
The direct scaling approach, however, contains a unique red ($c=3$) band at $\tilde\omega\approx 0.61$ and several green ($c=2$) bands around $\tilde\omega\approx0.58$, within the area filled with blue $c=2$ bands.
The red $c=3$ band seems to be degenerate with a blue $c=2$ band, which seems physically implausible and has been already observed by Ref.~\cite{Kozon2022Phys.Rev.Lett.} for a crystal structure with different pore sizes.
The more robustly grouped result of the clustering approach thus seem to be more physically plausible also in this case. 

The MBC approach offers another remarkable insight, namely, it assigns the green $c=1$ bands from the scaling approach to be $c=2$ bands.
Even by the visual inspection of energy density distribution $W(\vb r)$ it is hard to judge if the bands in question are $c=2$ or $c=0$ as concluded by the MBC and the k-means++ algorithm, respectively.
There does not seem to be a significant qualitative difference between the energy density distributions for the bands around $\tilde\omega\approx 0.65$ and the bands around $\tilde\omega\approx 0.67$, when visually compared.
All of them have $W(\vb r)$ extended throughout the whole supercell, but also with certain peaks at the defect positions.
Those peaks are, nevertheless, noticeably lower than the bands between $\tilde\omega\approx 0.52$ and $\tilde\omega\approx 0.60$.
The correct classification of the bands above $\tilde\omega\approx 0.65$ thus remains an open question at this time.

\begin{figure}[htbp]
\centering
    \includegraphics[width=\textwidth]{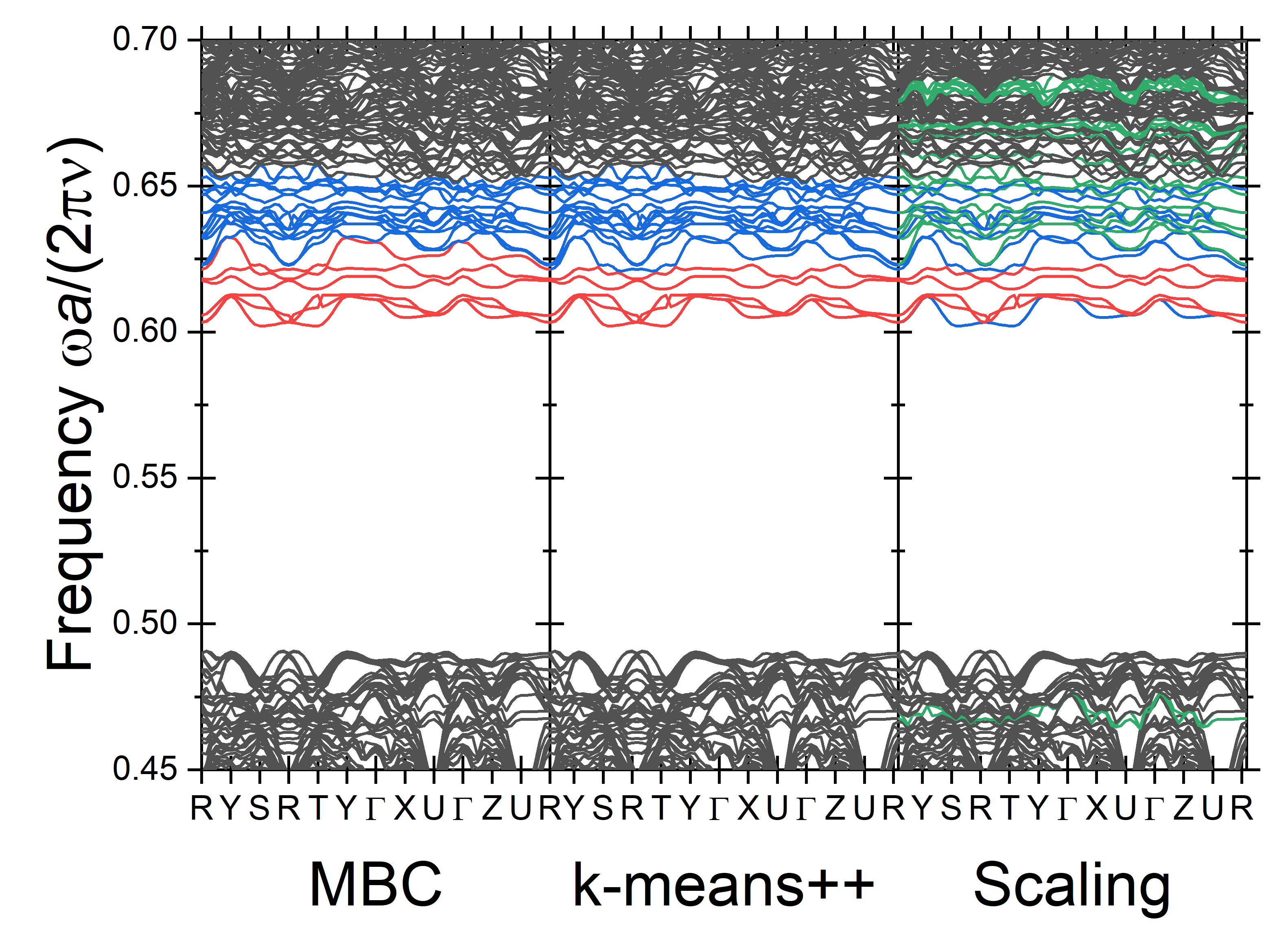}
  \caption{Band structure of an $N=3$ supercell of an inverse woodpile photonic crystal (see Appendix~\ref{iwpc}) with parameters $R=0.24a$, $R^\pr=0.8R$, with bands color-coded to indicate the confinement dimensionalities $c$ of each band. Red color corresponds to point-confined $c=3$, blue to line-confined $c=2$, green to plane-confined $c=1$ and black to extended $c=0$. The left panel shows the confinement classification for the MBC clustering, the central panel for the k-means++ clustering, and the right panel for the direct application of the scaling approach from Ref.~\cite{Kozon2022Phys.Rev.Lett.}.}
  \label{fig:N3r24rp08}
\end{figure}

Fig.~\ref{fig:N3r24rp08} shows the results of confinement identification on an $N=3$ supercell of the inverse woodpile photonic crystal with different structural parameters, namely the same unperturbed pore size $R=0.24a$, but increased defect pore size $R^\pr=0.8R$. 
The results of the clustering algorithms are almost identical, here, again avoiding unphysical cases.
The direct scaling approach finds two degenerate unphysical $c=1$ bands below the band gap at $\tilde\omega\approx 0.47$.
However, visual inspection of the energy density indicates that these two bands are unconfined ($c=0$), which was concluded by the clustering algorithms.
There are several other bands, between $\tilde\omega\approx 0.62$ and $\tilde\omega\approx 0.67$, that the scaling approach identifies as having $c=1$. 
Some of these bands are assigned to have $c=2$ (among the other blue bands), while others $c=0$ (above the other black bands) by the clustering algorithms, which seems a lot more consistent. 
Thus, also in this case, the clustering performs better than the direct scaling approach.

\begin{figure}[htbp]
\centering
    \includegraphics[width=\textwidth]{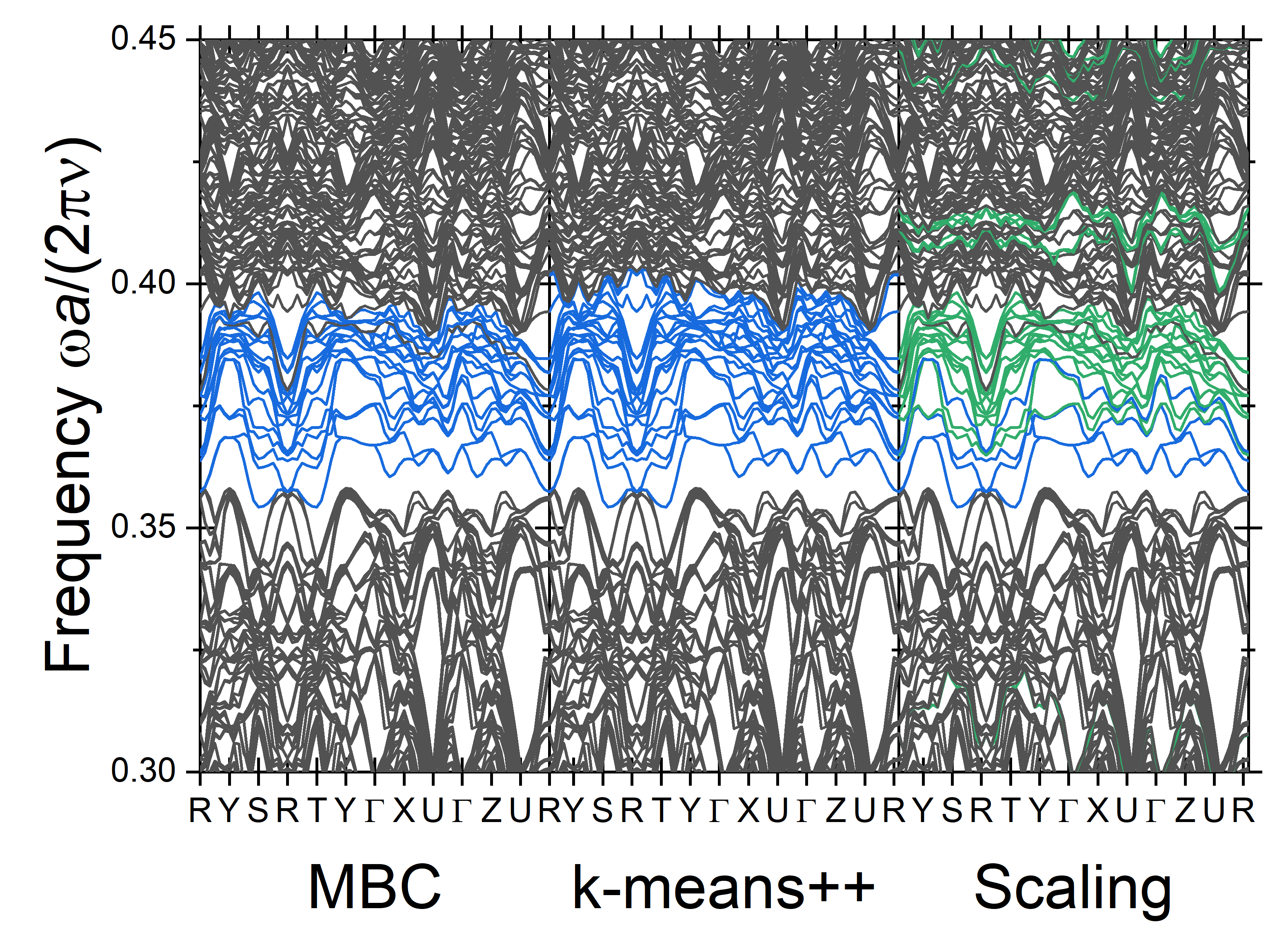}
  \caption{Band structure of an $N=3$ supercell of an inverse woodpile photonic crystal (see Appendix~\ref{iwpc}) with parameters $R=0.15a$, $R^\pr=0.5R$, with bands color-coded to indicate the confinement dimensionalities $c$ of each band. Red color corresponds to point-confined $c=3$, blue to line-confined $c=2$, green to plane-confined $c=1$ and black to extended $c=0$. The left panel shows the confinement classification for the MBC clustering, the central panel for the k-means++ clustering, and the right panel for the direct application of the scaling approach from Ref.~\cite{Kozon2022Phys.Rev.Lett.}.}
  \label{fig:N3r15rp05}
\end{figure}

Fig.~\ref{fig:N3r15rp05} shows the results of confinement identification on an $N=3$ supercell of the inverse woodpile photonic crystal with unperturbed pore size $R=0.15a$, and defect pore size $R^\pr=0.5R$. 
In this parameter set, the main pores are smaller than optimal, a situation typical of experiments~\cite{Leistikow2011Phys.Rev.Lett.,Adhikary2020Opt.Express,Uppu2021Phys.Rev.Lett.}.
In this case, the results obtained by both clustering algorithms almost coincide again.
The direct scaling approach again shows several unphysical green $c=1$ bands.
Some of these bands are classified by the clustering algorithms as $c=2$ and some as $c=0$.
Visual inspection of the energy density indicates that the $c=1$ bands around $\tilde\omega\approx0.41$, resulting from the direct scaling approach, are in fact $c=0$. 
This has been also identified by the clustering algorithms.
We thus again conclude that the clustering algorithms outperform the direct scaling approach.
The black $c=0$ band at $\tilde\omega=0.39$, resulting from the scaling and the MBC algorithm has similar energy-density distribution properties to its neighboring bands and therefore it looks like it should have been also labelled as $c=1$, which has been done only by the k-means++ algorithm.

\begin{figure}[htbp]
\centering
    \includegraphics[width=\textwidth]{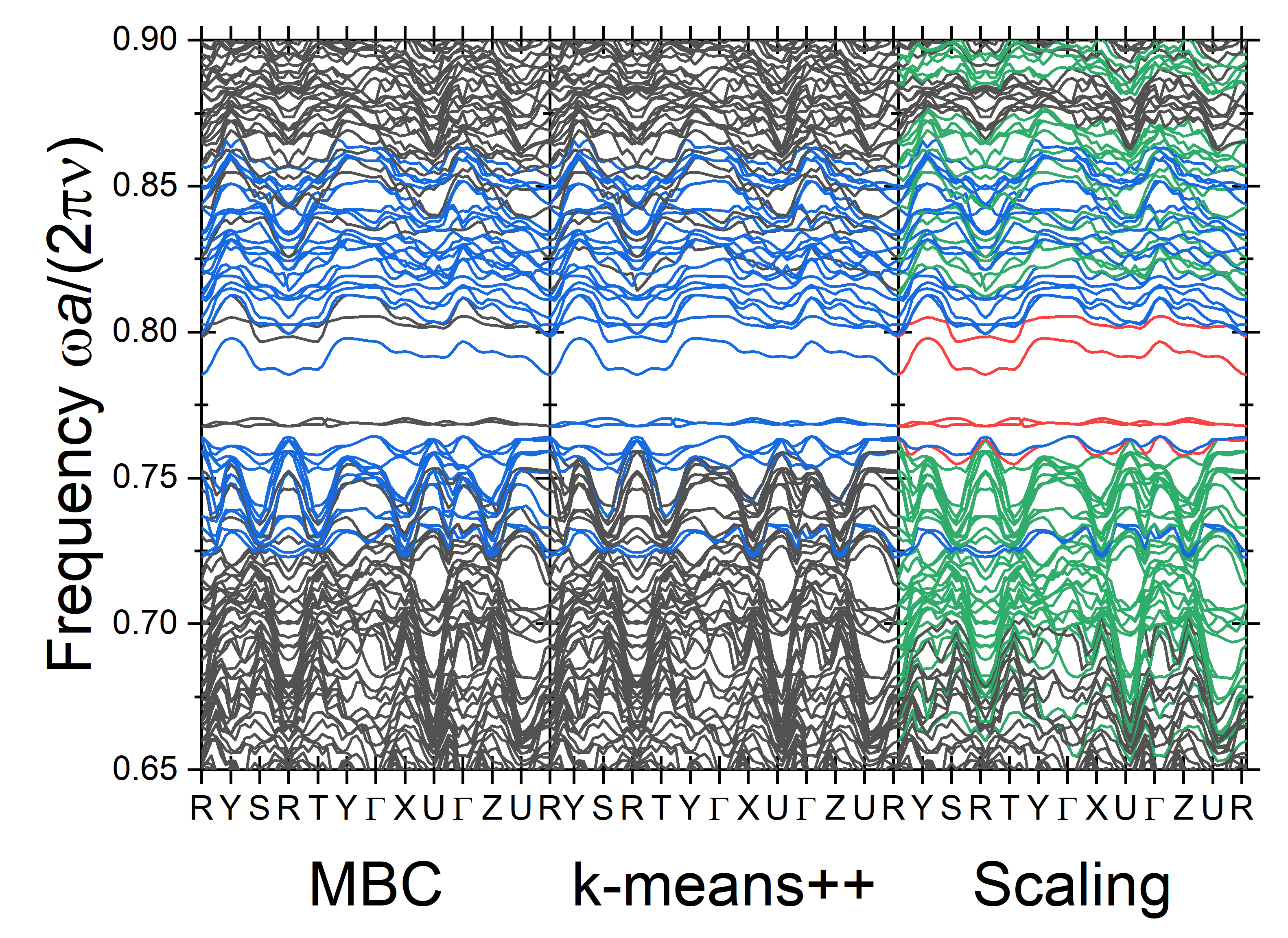}
  \caption{Band structure of an $N=3$ supercell of an inverse woodpile photonic crystal (see Appendix~\ref{iwpc}) with parameters $R=0.29a$, $R^\pr=0.5R$, with bands color-coded to indicate the confinement dimensionalities $c$ of each band. Red color corresponds to point-confined $c=3$, blue to line-confined $c=2$, green to plane-confined $c=1$ and black to extended $c=0$. The left panel shows the confinement classification for the MBC clustering, the central panel for the k-means++ clustering, and the right panel for the direct application of the scaling approach from Ref.~\cite{Kozon2022Phys.Rev.Lett.}.}
  \label{fig:N3r29rp05}
\end{figure}

Fig.~\ref{fig:N3r29rp05} shows the results of confinement identification on an $N=3$ supercell of the inverse woodpile photonic crystal with large unperturbed pores with $R=0.29a$, and defect pores with $R^\pr=0.5R$. 
This is a case where the direct scaling application overcomes the clustering.
Visual investigation of the energy-density distribution $W(\vb r)$ clearly shows that there are $c=3$ bands present in this system, such as those around $\tilde\omega\approx 0.77$.
The clustering algorithms with the help of DB* as a CVI thus clearly failed to identify the correct number of clusters here.

\begin{figure}[htbp]
\centering
    \includegraphics[width=\textwidth]{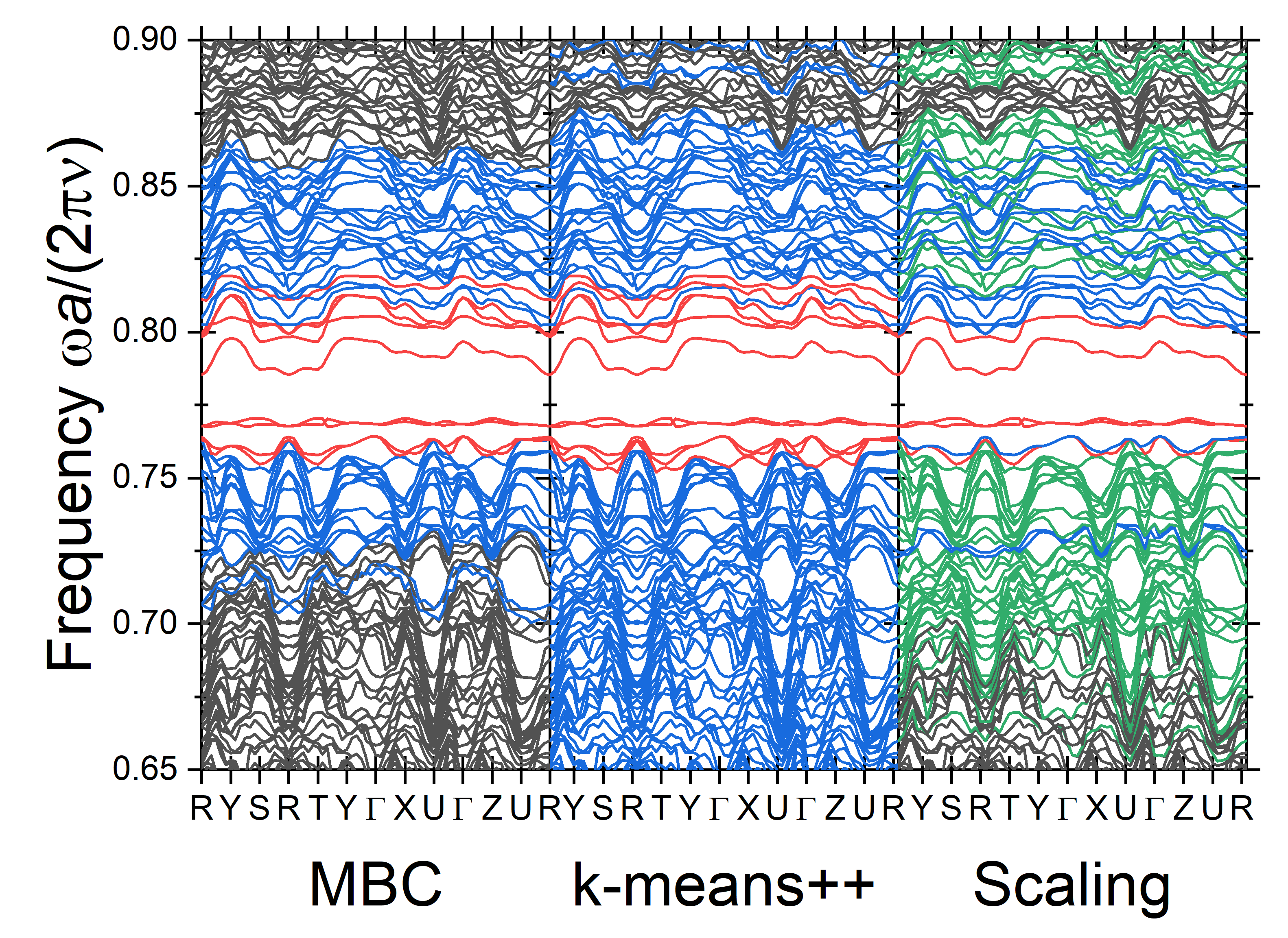}
  \caption{Band structure of an $N=3$ supercell of an inverse woodpile photonic crystal (see Appendix~\ref{iwpc}) with parameters $R=0.29a$, $R^\pr=0.5R$, with bands color-coded to indicate the confinement dimensionalities $c$ of each band. Red color corresponds to point-confined $c=3$, blue to line-confined $c=2$, green to plane-confined $c=1$ and black to extended $c=0$. The left panel shows the confinement classification for the MBC clustering, the central panel for the k-means++ clustering, and the right panel for the direct application of the scaling approach from Ref.~\cite{Kozon2022Phys.Rev.Lett.}. Here, we have explicitly chosen the number of cluster for the k-means++ algorithm and the clustering curves for the MBC, based on the direct scaling results.}
  \label{fig:N3r29rp05_corr}
\end{figure}

To exclude possible errors in CVI performance, we now eliminate the need for finding the correct number of clusters and confinement dimensionalities by explicitly imposing these values based on the results of the direct scaling. 
Fig.~\ref{fig:N3r29rp05_corr} shows the clustering classification where we have explicitly imposed the number of clusters $K=3$ for the k-means++ algorithm and the possible confinement dimensionalities $c=0,2,3$ for the MBC.
Now, both clustering algorithms identify $c=3$ bands similarly to the scaling algorithm.
Here, we see much better agreement regarding the $c=3$ bands among all three approaches.
The clustering algorithms obviously avoid the unphysical $c=1$ bands, thanks to the imposing of the number of clusters and the confinement dimensionalities before the analysis.
In this case, there is a remarkable difference between the line-confined $c=2$ bands identified by the k-means++ and the MBC algorithms.
The k-means++ algorithm finds a large number of these bands, especially also below the band gap of an unperturbed structure $\tilde\omega<0.72$, which seems physically highly implausible.
Indeed, visual inspection of the energy density of some of these bands indicates that they are $c=0$. 
Note that MBC also classifies a few bands below the gap as $c=2$.
To our surprise, the visual inspection of the energy density for some of these bands shows that, despite not being truly spatially confined, these bands do exhibit some confinement patterns at the defect cross-sections.
Therefore, MBC outperforms the other approaches here.

In this section, we investigated the confinement-classification performance of MBC, k-means++ and the direct scaling application methods on small supercells of various inverse woodpile structures.
The clustering algorithms clearly outperform the direct scaling approach if the implemented CVI identifies the correct number of clusters to use as the input to the algorithms. 
The performance of the two clustering algorithms is in most cases very similar.
Nevertheless, the MBC algorithm takes into account the underlying physics behind the clustering, which may prove beneficial in some cases, such as the one illustrated in Fig.~\ref{fig:N3r29rp05_corr}.
Moreover, MBC has an additional advantage of immediately assigning the clusters to their corresponding confinement dimensionalities, offering an additional advantage over the k-means++ algorithm where one has to assign the dimensionalities to the clusters manually.
Finally, to avoid errors in cases when the CVI fails, we find it valuable to first directly employ the scaling analysis of Ref.~\cite{Kozon2022Phys.Rev.Lett.} to identify the correct set of physically plausible confinement dimensionalities and then use those values as input for refinement by the MBC algorithm.


\section{Conclusion}

In this paper we investigate the application of clustering algorithms, which are particular techniques of unsupervised machine learning, to classify the confinement dimensionality of bands of confined states in photonic band gap superlattices.
We find that the use of clustering algorithms increases the accuracy of band confinement classification for small supercells. 
To overcome the lack of a ground truth, we employ cluster validity indices as a measure of partition correctness and as a way to find the correct number of clusters.
We analyze several CVIs and find that the most suitable validity measure for our application is the DB* variant of the Davies-Bouldin index.

We further propose two different algorithms to be used for band confinement identification: the k-means++ and our own model-based clustering (MBC).
We analyze the performance of these two algorithms in comparison to the direct application of scaling without clustering and find that the addition of clustering improves the accuracy of identification if the number of clusters is correctly found.
Nevertheless, we also discover that even with the use of the CVI, the clustering algorithms are not always able to find the correct number of clusters.
To overcome this issue, we propose to first directly apply scaling to find the number of physically valid clusters and then use that number as an input for the clustering algorithms to find the best band confinement classification.

Even though, as we have shown, both the k-means++ and our MBC algorithm usually perform well, in some cases the MBC algorithm is able to provide better results than the k-means++ algorithm.
Combined with the fact that MBC assigns confinement dimensionalities to the clusters inherently and automatically without any need for external input, we suggest the use of this algorithm over the k-means++ algorithm

In the future, it would be beneficial to devise a model-based CVI, measuring the cluster validity not only in terms of cohesion and separation but also with respect to the adherence to the scaling theory of confinement, in a similar manner as our model-based clustering algorithm works.
Furthermore, it would be interesting to explore the application of fuzzy clustering in order to include an error measure of assigning each point to its cluster~\cite{Jain1999ACMComput.Surv.}. 

\section*{Funding}
This research is supported by the NWO-CSER program, project “Understanding the absorption of interfering light for improved solar cell efficiency” under the Poject No. 680.93.14CSER035; the NWO-JCER program, project “Accurate and Efficient Computation of the Optical Properties of Nanostructures for Improved Photovoltaics” under the Project No. 680-91-084; the NWO-GROOT program, project “Self-Assembled Icosahedral Photonic Quasicrystals with a Band Gap for Visible Light” under the Project No. OCENW.GROOT.2019.071; the NWOTTW program P15-36 “Free-Form Scattering Optics” (FFSO); and the MESA+ Institute for Nanotechnology, section Applied Nanophotonics (ANP).

\section*{Disclosures}
The authors declare no conflicts of interest.

\appendix
\begin{subappendices}
\addappheadtotoc
\noindent{\huge\bfseries Appendices\par}

\section{The inverse woodpile structure}
\label{iwpc}
\begin{figure}[htbp]
	\includegraphics[width=\linewidth]{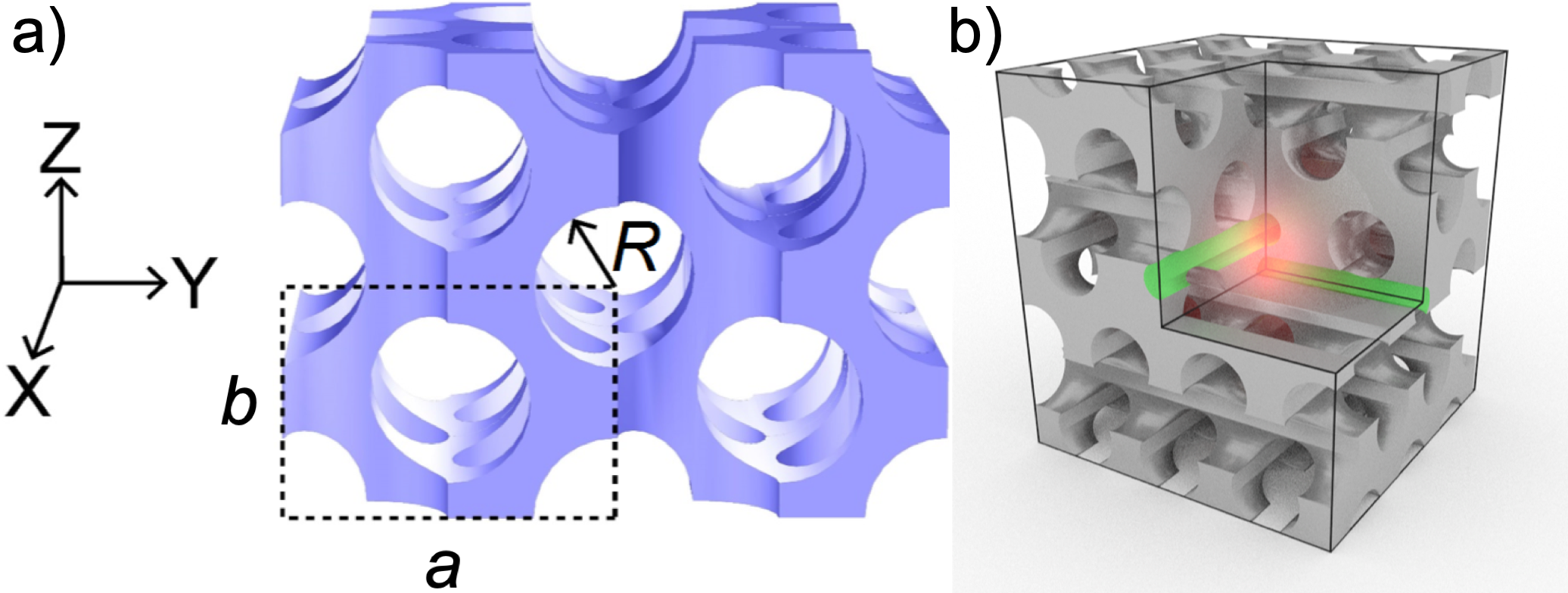}
	\centering
	\caption{(a) Structure of an unperturbed inverse woodpile photonic crystal. We use a tetragonal unit cell with lattice parameters $a$ in the $y$ direction and $b=a/\sqrt{2}$ in the $x$ and $z$ directions. The pore radius is denoted by $R$. The figure shows an $N=2$ supercell, with one unit cell designated by the dashed border. (b) Design of the defect. The radius $R^\pr$ of two proximate defect pores (shown in green) is altered. At their intersection a region with excess of one material is created that serves as a point defect (red glow).}
	\label{fig:IWPC}
\end{figure}

The 3D inverse woodpile photonic band gap crystal consists of bulk silicon ($\eps = 12.1$, see Ref.~\cite{Hillebrand2003J.Appl.Phys.}), in which nanopores of radius $R$ filled with air ($\eps=1$) are etched \cite{Ho1994SolidStateCommun.,Leistikow2011Phys.Rev.Lett.}.
We model the unperturbed crystal using a tetragonal unit cell with lattice parameters $a$ in the $y$ direction and $b=a/\sqrt{2}$ in the $x$ and $z$ directions.
This unperturbed structure is depicted in Fig.~\ref{fig:IWPC}(a).

A defect can be incorporated in the structure by altering the radius $R^\pr$ of two proximate nanopores, as illustrated in Fig.~\ref{fig:IWPC}(b)~\cite{Woldering2014Phys.Rev.B}.
This results in the creation of a region with excess of either air ($R^\pr>R$) or silicon ($R^\pr<R$) at the intersection of the two defect pores, serving as a point defect.
We include one defect of this type per supercell.

\section{Overview of the studied CVIs}
\label{CVIs}
Here, the mathematical definition for each CVI is provided along with some intuitive explanation of each formula, mainly in terms of cluster cohesion and cluster separation.
For a comprehensive review, see Ref.~\cite{Arbelaitz2013PatternRecognition}.
We denote the centroid of the dataset $\XX$ containing $M$ data points as $\vb  X= \frac{1}{M}\sum_{\vb x_i\in\XX} x_i$.

\subsection{Silhouette}

The Silhouette index~\cite{Rousseeuw1987JournalofComputationalandAppliedMathematics} is defined as
\begin{equation}
    \mr{Sil}(\CCC) \eqdef \frac{1}{M} \sum_{C_i \in \CCC} \sum_{\vb x_j \in C_i} \frac{b(\vb x_j,C_i) - a(\vb x_j, C_i)}{\max\{a(\vb x_j, C_i),b(\vb x_j, C_i)\}}
\end{equation}
and is aimed to be maximized for the best clustering outcome.
Here, 
\begin{equation}
    a(\vb x_i,C_k) \eqdef\frac{1}{M_i} \sum_{\vb x_j \in C_k} \norm{\vb x_j - \vb x_i}^2,\quad \text{for } \vb x_i \in C_k,
\end{equation}
represents the cohesion of the cluster $C_k$ given as a distance between a chosen point $\vb x_i \in C_k$ and all other points in the given cluster.
Clearly, the smaller $a$ is, the better the cluster cohesion. 
Furthermore, the function
\begin{equation}
    b(\vb x_i,C_k) \eqdef\min_{C_l \in \CCC \setminus C_k} \left\{\frac{1}{M_l} \sum_{\vb x_j \in C_l} \norm{\vb x_j - \vb x_l}^2 \right\} ,\quad \text{for } \vb x_i \in C_k,
\end{equation}
represents the separation of the cluster in terms of nearest neighbor distance from the other clusters $C_l, l\neq k$ to the point $\vb x_i \in C_k$.
The larger $b$ is, the further apart different clusters are from each other and thus the better the cluster separation is.

\subsection{Calinski-Harabasz} 
The Calinski-Harabasz index~\cite{Calinski1974Comm.inStats.-Theory&Methods} is defined as
\begin{equation}\label{calinski}
    \mr{CH}(\CCC) \eqdef \frac{M - K}{K - 1}  \frac{\sum_{C_i \in \CCC} M_i \norm{ \vb c_i - \vb X}^2}{\sum_{C_i \in \CCC} \sum_{\vb x_j \in C_i } \norm{ \vb x_j - \vb c_i}^2}.
\end{equation}
A high value of CH is assumed to correspond to better data clustering.

Intuitively, the distance between the global centroid $\vb X$ and the cluster centroids $\vb c_i$ in the numerator acts as a cluster separation measure, with larger number corresponding to better separation. 
The cluster cohesion is represented by the denominator as the sum of distances between the cluster centroids and their constituent data points.
Smaller distances between the cluster centroid and the cluster data obviously correspond to a more cohesive cluster. 
For a good clustering, one thus aims for maximizing the numerator and minimizing the denominator of CH, thus maximizing its overall value.

\subsection{Davies-Bouldin}
The Davies-Bouldin index~\cite{Davies1979IEEETrans.PatternAnal.Mach.Intell.} is defined as 
\begin{equation}
    \mr{DB}(\CCC) \eqdef \frac{1}{K} \sum_{C_i \in \CCC} \max_{C_j \in \CCC \setminus C_i} \left\{ \frac{S(C_i)+S(C_j)}{\norm{\vb c_i - \vb c_j}^2}\right\},
\end{equation}
where
\begin{equation}
    S(C_i) \eqdef \frac{1}{M_i} \sum_{\vb x_j \in C_i} \norm{\vb x_j - \vb c_i}^2.
\end{equation}
DB is expected to decrease with better partioning.

Here, the cohesion of the cluster $C_i$ is gauged by the function $S(C_i)$ in the numerator, as the sum of the intracluster distances, \ie , the sum of the distance of each data point assigned to the cluster $C_i$ to the cluster centroid.
The separation is measured in the denominator simply as the distance between the cluster centroids.
A good clustering result will have a small numerator and a large denominator, thus resulting in a small value of DB.

\subsection{Davies-Bouldin*}
This variant on the Davies-Bouldin algorithm, described in Ref.~\cite{Kim2005PatternRecognitionLetters}, is defined as
\begin{equation}
    \mr{DB^*}(\CCC) \eqdef \frac{1}{K} \sum_{C_i \in \CCC} \frac{\max_{C_j \in \CCC \setminus C_i} \left\{ S(C_i)+S(C_j)\right\}}{\min_{C_j \in \CCC \setminus C_i} \left\{ \norm{\vb c_i - \vb c_j}^2\right\}}
\end{equation}
and is expected to attain low values for good clustering outcomes.
The fact that the standard DB minimizes the maximum of the ratio of cluster cohesion and cluster separation can lead to pathological cases, as described by Ref.~\cite{Kim2005PatternRecognitionLetters}.
To remedy this, DB* instead maximizes the cluster cohesion and minimizes the cluster separation independently.

\subsection{COP}
The COP index~\cite{Gurrutxaga2010PatternRecognition} is defined as
\begin{equation}
    COP(\CCC) \eqdef \frac{1}{M} \sum_{C_i \in \CCC} M_i \frac{ \frac{1}{M_i} \sum_{\vb x_j \in C_i} \norm{\vb x_j -\vb c_i}^2}{\min_{\vb x_j \notin C_i} \max_{\vb x_k \in C_i} \left\{\norm{ \vb x_j - \vb x_k}^2\right\}}
\end{equation}
and is expected to be small for good clustering results.
Here, the cluster cohesion is measured in the numerator as the average of intracluster distances and the cluster separation is measured by minimizing the furthest-neighbor distance to a given cluster in the denominator.
One aims to minimize the numerator and maximize the denominator, thus maximizing the value of COP.
We note here that, due to the furthest-neighbor distance measure, long and thin cluster can possibly skew this measure by overestimating the actual separation between the clusters.

\subsection{S\_Dbw index}
The S\_Dbw index~\cite{Halkidi2001Proc.2001IEEEInt.Conf.DataMin.} is defined as
\begin{equation}
\begin{aligned}
    \mr{S\_Dbw}(\CCC) \eqdef&\frac{1}{K} \sum_{C_i \in \CCC} \frac{\norm{ \sigma(C_i)}}{\norm{\sigma(X)}} \\
    &+ \frac{1}{K(K-1)} \sum_{C_i \in \CCC} \sum_{C_j \in \CCC \setminus  C_i } \frac{\rho(C_i, C_j)}{\max\left\{\rho(C_i), \rho(C_j)\right\}}
    \end{aligned}
    \label{SDbw}
\end{equation}
and is expected to attain small values for good partitions.
Here, $\boldsymbol \sigma (\XX) $ is the variance of each component of the data set $\XX$.
Its $p$-th component is defined as
\begin{equation}
    \sigma^p (\XX) \eqdef \frac{1}{|X|} \sum_{\vb x_j \in \XX} (x_j^p - X^p)^2,
\end{equation}
Furthermore, $\tau(\CCC)$ is the standard deviation of the partition $\CCC$, defined as
\begin{equation}
     \tau(\CCC) \eqdef \frac{1}{K} \sqrt{\sum_{C_i \in \CCC} \norm{\boldsymbol \sigma(C_i)}}.
\end{equation}
We further define the terms
\begin{equation}
   \rho(C_i) \eqdef \sum_{\vb x_j \in C_i} \theta(\vb x_j,\vb c_i),
\end{equation}
and
\begin{equation}
    \rho(C_i, C_j) \eqdef \sum_{\vb x_k \in C_i \cup C_j}\theta \left( \vb x_k, \frac{\vb c_i + \vb c_j}{2} \right),
\end{equation}
where
\begin{equation}
      \theta(\vb x_j, C_i)\eqdef\begin{cases}0~&{\text{ if }}~\norm{\vb x_j - \vb c_i} > \tau(\CCC)~\\1~&{\text{ else}}\end{cases}.
\end{equation}
The term $\rho(C_i)$ evaluates the number of points in a cluster $C_i$ within the standard deviation of the partition and the term $\rho(C_i, C_j)$ evaluates the number of points at the midpoint between the centers of $C_i$ and $C_j$.
A good clustering corresponds to large $\rho(C_i), \rho(C_j)$ and small $\rho(C_i, C_j)$ for each $i,j\le K, i\ne j$, which translates into a small value of the second term in~\eq{SDbw}.
The first term in~\eq{SDbw} is termed intracluster variance and is a measure of the cluster cohesion in terms of a mean cluster variance. 

The first term in Eq.~\eq{SDbw} thus corresponds to the cluster cohesion and the second term represents the cluster separation~\cite{Halkidi2001Proc.2001IEEEInt.Conf.DataMin.}.
Unlike all the other CVIs described in this paper, S\_Dbw relates these two terms by addition, instead of a ratio.
Overall S\_Dbw aims to minimize both of these terms for a good clustering result.

\end{subappendices}


\bibliography{Clustering_algorithms_for_confinement}

\end{document}